\documentclass{tufte-handout}
\usepackage[utf8]{inputenc}
\usepackage{graphicx}
\usepackage{amsmath}
\usepackage{cleveref}
\usepackage{subfig}
\usepackage{color}
\usepackage[svgnames]{xcolor}
\usepackage{rotating}
\usepackage{DejaVuSansMono}
\usepackage[normalem]{ulem} 
\usepackage{multirow}

\crefname{figure}{Fig.}{Figs.}
\Crefname{figure}{Figure}{Figures}

\title[Microscopic Epidemic Model and MARL]{A Microscopic Epidemic Model and Pandemic Prediction Using Multi-Agent Reinforcement Learning}
\author{Changliu Liu\thanks{C. Liu is with Carnegie Mellon University, Pittsburgh PA 15213. Email: \tt cliu6@andrew.cmu.edu}}
\date{ }

\begin{document}

\maketitle

\begin{abstract}
This paper introduces a microscopic approach to model epidemics, which can explicitly consider the consequences of individual's decisions on the spread of the disease. We first formulate a microscopic multi-agent epidemic model where every agent can choose its activity level that affects the spread of the disease. Then by minimizing agents' cost functions, we solve for the optimal decisions for individual agents in the framework of game theory and multi-agent reinforcement learning. Given the optimal decisions of all agents, we can make predictions about the spread of the disease. We show that there are negative externalities in the sense that  infected agents do not have enough incentives to protect others, which then necessitates external interventions to regulate agents' behaviors. 
In the discussion section, future directions are pointed out to make the model more realistic.\sidenote{This paper is adapted from the lecture notes for Lectures 20 of 16-899 Adaptive Control and Reinforcement Learning (Spring 2020) at CMU. The course slides are available at \url{https://piazza.com/class_profile/get_resource/k54pll5057h79m/k8qaky3o8x65bm}. The code is available at \url{https://github.com/intelligent-control-lab/Microscopic_Epidemic_Model}.}
\end{abstract}

\section{Introduction}
With the COVID-19 pandemic souring across the world, a reliable model is needed to describe the observed spread of the disease, make predictions about future, and guide public policy design to control the spread. 

\paragraph{Existing Epidemic Models} There are many existing macroscopic epidemic models.\cite{daley2001epidemic} For example, the SI model describes the growth of infection rate as the product of the current infection rate and the current susceptible rate. The SIR model further incorporates the effect of recovery into the model, i.e., when the infected population turns into immune population after a certain period of time. The SIRS model considers the case that immunity is not for lifetime and that the immune population can become susceptible population again. In addition to these models, the SEIR model incorporates the incubation period into analysis. Incubation period refers to the duration before symptoms show up.\cite{yan2006seir} The most important factor in all those models is $R0$, the regeneration number, which tells how fast the disease can spread. $R0$ can be regressed from data.

\paragraph{Limitations of Existing Models} Although these models are useful in predicting the spread of epidemics, they lack the granularity needed for analyzing individual behaviors during an epidemic and understanding the relationship between individual decisions and the spread of the disease.\cite{barrett2009estimating} For example, many countries now announced ``lock-down", ``shelter-in-place", ``stay-at-home", or similar orders. However, their effects are very different across different countries, or even across different counties in the same country. One factor that can possibly explain these differences is the cultural difference. In different cultures, individuals make different choices. For instance, in the west, people exhibit greater inertia to give up their working/life routines so that they do not follow the orders seriously. While in the east, people tend to obey the rules better. These different individual choices can result in significantly different outcomes in disease propagation that cannot be captured by a macroscopic model.


\paragraph{A Microscopic Epidemic Model} In this paper, we develop a microscopic epidemic model by explicitly considering individual decisions and the interaction among different individuals in the population, in the framework of multi-agent systems. The aforementioned cultural difference can be understood as a difference in agents' cost functions, which then affect their behaviors when they are trying to minimize their cost functions. The details of the microscopic epidemic model will be explained in the next section, followed by the analysis of the dynamics of the multi-agent system, and the prediction of system trajectories using multi-agent reinforcement learning. The model is still in its preliminary form. In the discussion section, future directions are pointed out to make the model more realistic.

\section{Microscopic Epidemic Model}
Suppose there are $M$ agents in the environment. Initially, $m_0$ agents are infected. Agents are indexed from $1$ to $M$. Every agent has its own state and control input. The model is in discrete time. The time interval is set to be one day. The evolution of the infection rate for consecutive days depends on agents' actions. The questions of interest are: How many agents will eventually be infected? How fast they will be infected? How can we slow down the growth of the infection rate?

\subsection{Agent Model}
We consider two state values for an agent, e.g., for agent $i$, $x_i = 0$ means healthy (susceptible), $x_i = 1$ means infected.
Everyday, every agent $i$ decides its level of activities $u_i\in[0,1]$. The level of activities for agent $i$ can be understood as the expected percentage of other agents in the system that agent $i$ wants to meet. For example, $u_i = 1/M$ means agent $i$ expects to meet one other agent. The actual number of agents that agent $i$ meets depends not only on agent $i$'s activity level, but also on other agents' activity level. For example, if all other agents choose an activity level $0$, then agent $i$ will not be able to meet any other agent no matter what $u_i$ it chooses. 
Mathematically, the chance for agent $i$ and agent $j$ to meet each other depends on the minimum of the activity levels of these two agents, i.e., $\min\{u_i, u_j\}$. In the extreme cases, if agent $i$ decides to meet everyone in the system by choosing $u_i = 1$, then the chance for agent $j$ to meet with agent $i$ is $u_j$. If agent $i$ decides to not meet anyone in the system by choosing $u_i = 0$, then the chance for agent $j$ to meet with agent $i$ is $0$.

Before we derive the system dynamic model, the assumptions are listed below:\sidenote{These assumptions can all be relaxed in future work. They are introduced mainly for the simplicity of the discussion.}
\begin{enumerate}
\item In the agent model, we only consider two states: healthy (susceptible) and infected. All healthy agents are susceptible to the disease. There is no recovery and no death for infected agents. There is no incubation period for infected agents, i.e., once infected, the agent can start to infect other healthy agents. To relax this assumption, we may introduce more states for every agent.
\item The interactions among agents are assumed to be uniform, although it is not true in the real world. In the real world, given a fixed activity level, agents are more likely to meet with close families, friends, colleagues than strangers on the street. To incorporate this non-uniformity into the model, we need to redefine the chance for agent $i$ and agent $j$ to meet each other to be $\beta_{i,j}\min\{u_i, u_j\}$, where $\beta_{i,j}\in[0,1]$ is a coefficient that encodes the proximity between agent $i$ and agent $j$ and will affect the chance for them to meet with each other. For simplicity, we assume that the interaction patterns are uniform in this paper.
\item Meeting with infected agents will result in immediate infection. To relax this assumption, we may introduce an infection probability to describe how likely it is for a healthy agent to be infected if it meets with an infected agent.
\end{enumerate}

\subsection{System Dynamic Model}
On day $k$, denote agent $i$'s state and control as $x_{i,k}\in\mathcal{X}$ and $u_{i,k}\in\mathcal{U}$. By definition, the agent state space is $\mathcal{X} = \{0,1\}$ and the agent control space is $\mathcal{U}=[0,1]$. The system state space is denoted $\mathcal{X}^M:= \mathcal{X}\times\cdots \times \mathcal{X}$. The system control space is denoted $\mathcal{U}^M:=\mathcal{U}\times\cdots \times \mathcal{U}$. Define $m_k = \sum_{i} x_{i,k}$ as the number of infected agents at time $k$. The set of infected agents is denoted:
\begin{equation}
\mathcal{I}_k := \{i: x_{i,k}= 1\}.
\end{equation}

The state transition probability for the multi-agent system is a mapping
\begin{equation}
\mathbb{T}: \mathcal{X}^M \times \mathcal{U}^M \times \mathcal{X}^M \mapsto [0,1].
\end{equation}

According to the assumptions, an infected agent will always remain infected. Hence the state transition probability for an infected agent $i$ does not depend on other agents' states or any control. However, the state transition probability for a healthy agent $i$ depends on others. The chance for a healthy agent $i$ to not meet an infected agent $j\in\mathcal{I}_k$ is $1-\min\{u_i,u_j\}$. A healthy agent can stay healthy if and only if it does not meet any infected agent, the probability of which is $\Pi_{j\in\mathcal{I}_k} (1-\min\{u_i,u_j\})$. Then the probability for a healthy agent to be infected is $1-\Pi_{j\in\mathcal{I}_k} (1-\min\{u_i,u_j\})$. From the expression $\Pi_{j\in\mathcal{I}_k} (1-\min\{u_i,u_j\})$, we can infer that: the chance for a healthy agent $i$ to stay health is higher if
\begin{itemize}
\item the agent $i$ limits its own activity by choosing a smaller $u_i$;
\item the number of infected agents is smaller; 
\item the infected agents in $\mathcal{I}_k$ limit their activities.
\end{itemize}

The state transition probability for an agent $i$ is summarized in \cref{table:dynamics}. 
\begin{table}[h]
\caption{The state transition probability from $x_{i,k}$ to $x_{i,k+1}$ for an agent.}
\begin{center}\small{
\begin{tabular}{c|cc}
& $x_{i,k}=0$ & $x_{i,k}=1$\\
\hline
 $x_{i,k+1}=0$ & $\Pi_{j\in\mathcal{I}_k} (1-\min\{u_i,u_j\})$ & $0$\\
 $x_{i,k+1} = 1$& $1-\Pi_{j\in\mathcal{I}_k} (1-\min\{u_i,u_j\})$ & $1$
\end{tabular}}
\end{center}
\label{table:dynamics}
\end{table}%

\begin{marginfigure}
\begin{center}
\includegraphics[width=3.5cm]{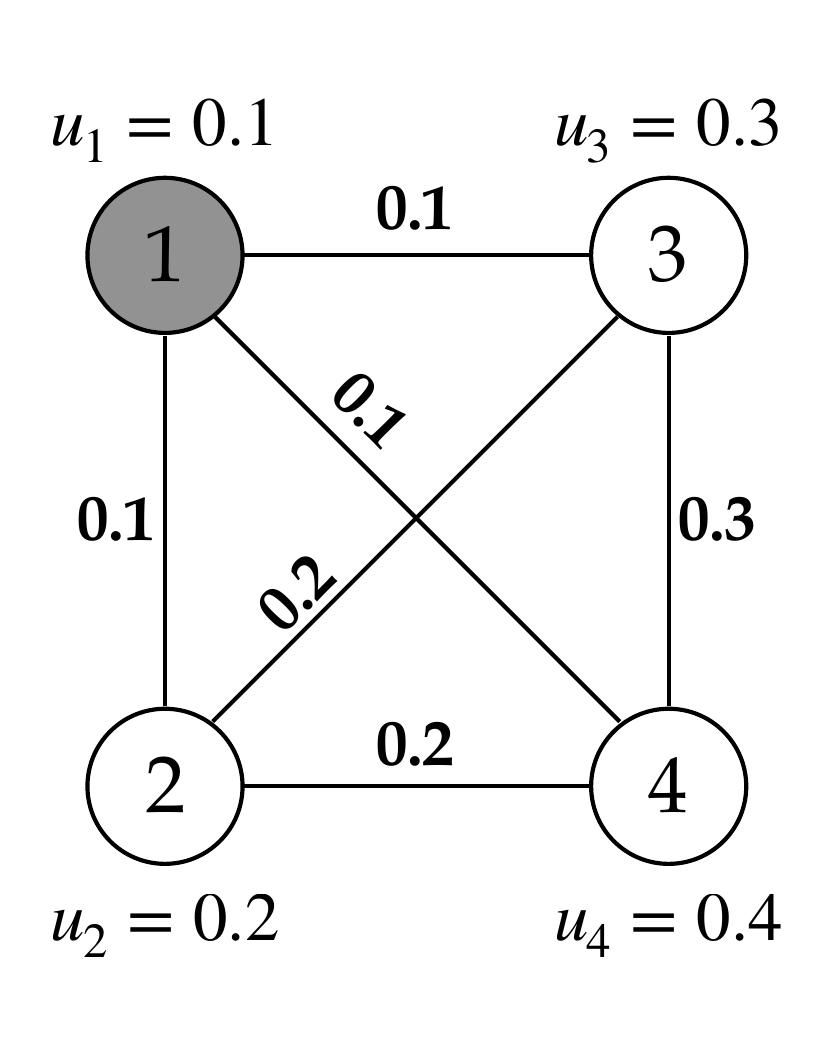}
\caption{Example of a four-agent system. Agent $1$ is infected (shaded). Other agents are healthy. The numbers on the links denote the probability for agents to meet with each other, which depend on the chosen activity levels of different agents.}
\label{fig:example}
\end{center}
\end{marginfigure}

\paragraph{Example} 
Consider a four-agent system shown in \cref{fig:example}. Only agent~$1$ is infected. And the agents choose the following activity levels: $u_1 = 0.1, u_2 = 0.2, u_3 = 0.3, u_4 = 0.4$. Then the chance $p_{i,j}$ for agents $i$ and $j$ to meet with each other is $p_{1,2} = p_{1,3} = p_{1,4} = 0.1$, $p_{2,3} = p_{2,4} = 0.2$, and $p_{3,4} = 0.3$. Note that $p_{i,j} = p_{j,i}$. The chance for agents $2$, $3$, and $4$ to stay healthy is $0.9$, although they have different activity levels.

\subsection{Case Study}
Before we start to derive the optimal strategies for individual agents and analyze the closed-loop multi-agent system, we first characterize the (open-loop) multi-agent system dynamics by Monte Carlo simulation according to the state transition probability in \cref{table:dynamics}.

Suppose we have $M=1000$ agents. At the beginning, only agent~$1$ is infected. We consider two levels of activities: normal activity level $u$ and reduced activity level $u^*$. The two activity levels are assigned to different agents following different strategies as described below. In particular, we consider ``no intervention" case where all agents continue to follow the normal activity level, ``immediate isolation" case where the activity levels of infected agents immediately drop to the reduced level, ``delayed isolation" case where the activity levels of infected agents drop to the reduced level after several days, and ``lockdown" case where the activity levels of all agents drop to the reduced level immediately.
\begin{marginfigure}
\begin{center}
\includegraphics[width=\linewidth]{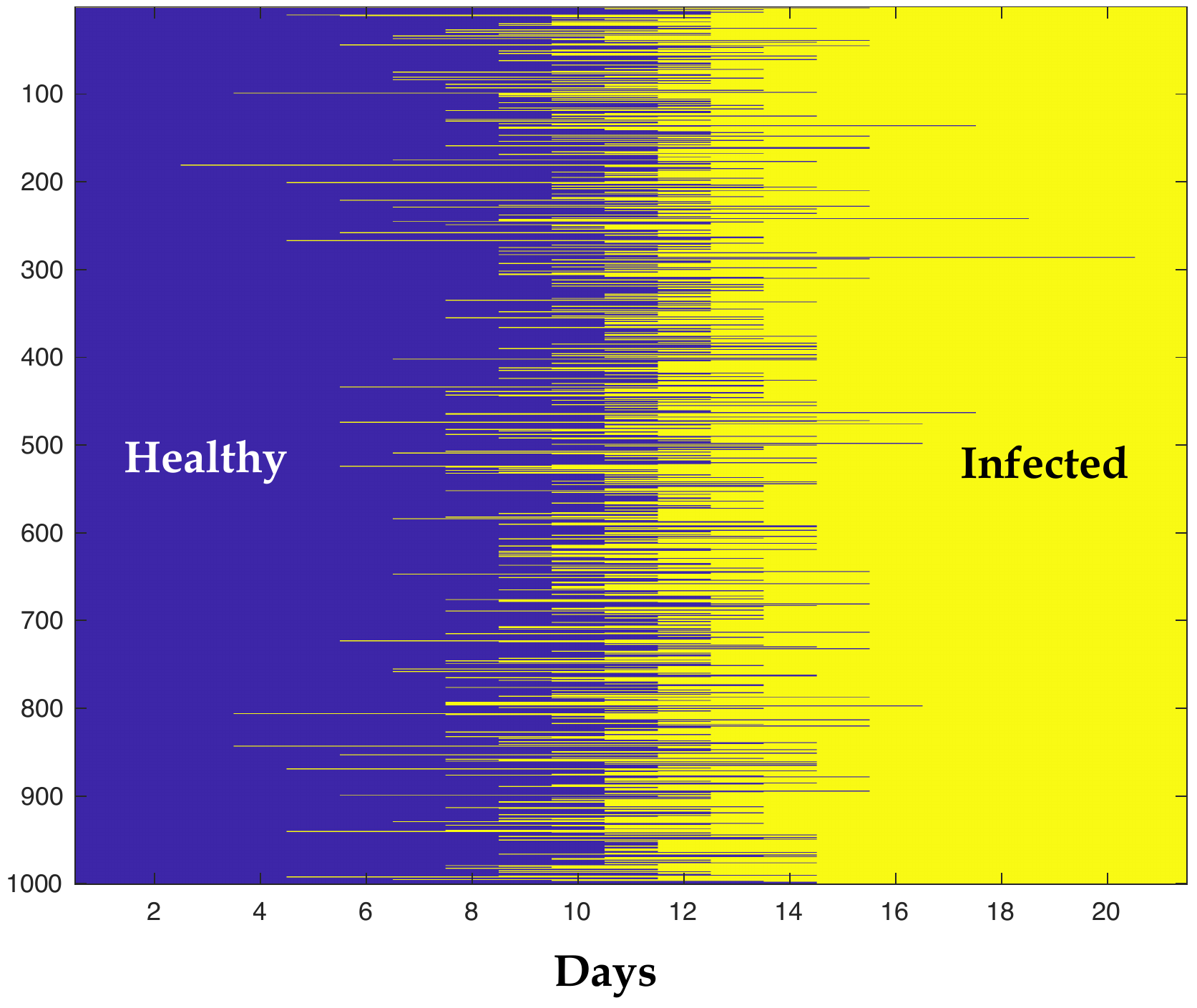}
\caption{Illustration of the result of one Monte Carlo simulation when all agents have the activity level $u = 1/M$. The horizontal axis corresponds to day $k$. The vertical axis corresponds to agent ID $i$. The color in the graph represents the value of $x_{i,k}$, blue for $0$ (healthy) and yellow for $1$ (infected).}
\label{fig:monte_carlo}
\end{center}
\end{marginfigure}

\begin{figure*}[htbp]
\begin{center}
\includegraphics[width = \linewidth]{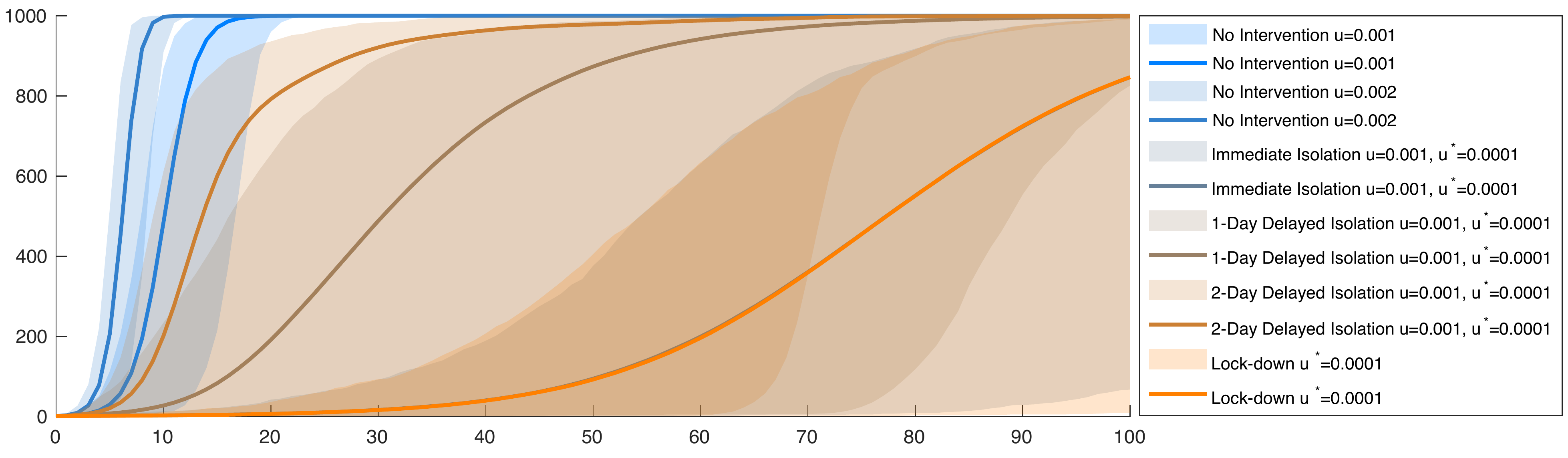}
\caption{The growth of the infection rate under different activity levels of agents. For every scenario, the result is extracted from 200 Monte Carlo simulations. The horizontal axis corresponds to day $k$. The vertical axis corresponds number of infected agents $m_k$. The solid curves are the average $m_k$ and the shaded area corresponds to the range from min $m_k$ to max $m_k$.}
\label{fig:open_loop}
\end{center}
\end{figure*}

For each case, we simulate 200 system trajectories and compute the average, maximum, and minimum $m_k$ (number of infected agents) versus $k$ from all trajectories. A system trajectory in the ``no intervention" case is illustrated in \cref{fig:monte_carlo}, where $u=1/M$ for all agents. The $m_k$ trajectories under different cases are shown in \cref{fig:open_loop}, where the solid curves illustrate the average $m_k$ and the shaded area corresponds to the range from min $m_k$ to max $m_k$. The results are explained below.

\begin{itemize}
\item Case 0: no intervention.

All agents keep the normal activity level $u$. The scenarios for $u= 1/M$ and $u= 2/M$ are illustrated in \cref{fig:open_loop}. As expected, a higher activity level for all agents will lead to faster infection. The trajectory of $m_k$ has a $S$ shape, whose growth rate is relatively slow when either the infected population is small or the healthy population is small, and is maximized when $50\%$ agents are infected. It will be shown in the following discussion that (empirical) macroscopic models also generate $S$-curves. 
 
\item Case 1: immediate isolation of infected agents.

The activity levels of infected agents immediately drop to $u^*$, while others remain $u$. The scenario for $u = 1/M$ and $u^* = 0.1/M$ is illustrated in \cref{fig:open_loop}. Immediate isolation significantly slows down the growth of the infections rate. As expected, it has the best performance in terms of flattening the curve, same as the lockdown case. The trajectory also has a $S$ shape.
 
\item Case 2: delayed isolation of infected agents.

The activity levels of infected agents drop to $u^*$ after $T$ days, while others remain $u$. In the simulation, $u = 1/M$ and $u^* = 0.1/M$.  The scenarios for $T=1$ and $T=2$ are illustrated in \cref{fig:open_loop}. As expected, the longer the delay, the faster the infection rate grows, though the growth of the infection rate is still slower than the ``no intervention" case. Moreover, the peak growth rate (when $50\%$ agents are infected) is higher when the delay is longer.

\item Case 3: lockdown.

The activity levels of all agents drop to $u^*$. The scenario for $u^* = 0.1/M$ is illustrated in \cref{fig:open_loop}. As expected, it has the best performance in terms of flattening the curve, same as the immediate isolation case.\sidenote{In the case that infected population can be asymptomatic or have a long incubation period before they show any symptom, like what we observe for COVID-19, immediate identification of infected person and then immediate isolation is not achievable. Then lockdown is the only best way to control the spread of the disease in our model.}
\end{itemize}

Since the epidemic model is monotone, every agent will eventually be infected as long as the probability to meet infected agents does not drop to zero.
Moreover, we have not discussed decision making by individual agents yet. The activity levels are just predefined in the simulation.

\paragraph{Remark}
The model we introduced is microscopic, in the sense that interactions among individual agents are considered. The simulated open-loop trajectories are indeed similar to those from a macroscopic model. Since only susceptible and infected populations are considered in the proposed microscopic model, we then compare it with the macroscopic Susceptible-Infected (SI) model. 
Define the state $s\in[0,1]$ as the fraction of infected population. The growth of the infected population is proportional to the susceptible population and the infected population. 
Suppose the infection coefficient is $\beta$, the system dynamics in the SI model follow:
\begin{equation}
\dot s = \beta s (1-s).
\end{equation}
\begin{marginfigure}
\begin{center}
\includegraphics[width=\linewidth]{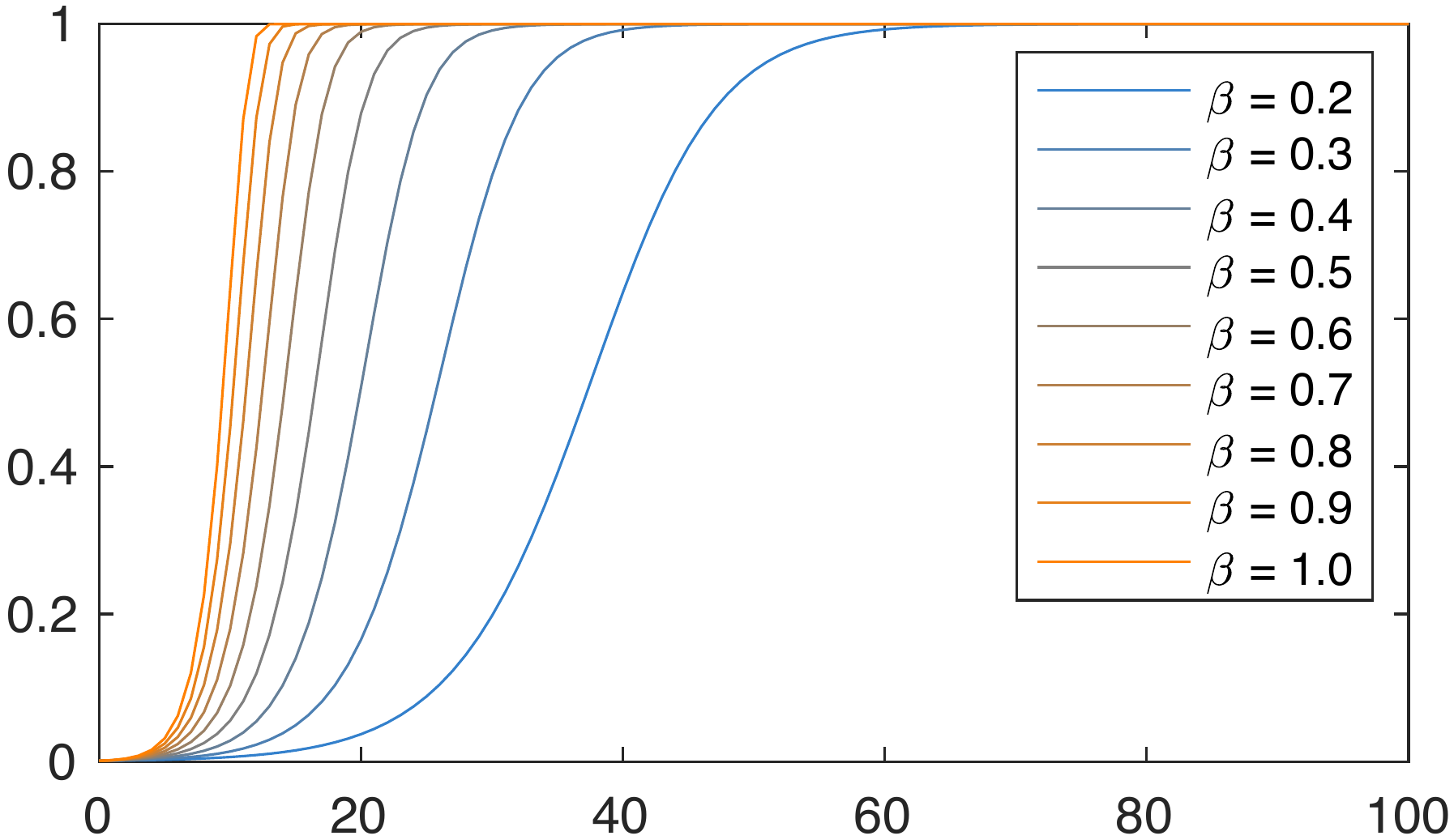}
\caption{The system trajectories in the macroscopic SI model. The horizontal axis corresponds to days. The vertical axis corresponds to the infection rate $s$.}
\label{fig:SI}
\end{center}
\end{marginfigure}
We simulate the system trajectory under different infection coefficients as shown in \cref{fig:SI}. The trajectories also have S shapes, similar to the ones in the microscopic model. However, since this macroscopic SI model is deterministic, there is no ``uncertainty" range as shown in the microscopic model. The infection coefficient $\beta$ depends on the agents' choices of activity levels. However, there is not an explicit relationship yet. It is better to directly use the microscopic model to analyze the consequences of individual agents' choices.

\section{Distributed Optimal Control}
This section tries to answer the following question: in the microscopic multi-agent epidemic model, what is the best control strategy for individual agents?
To answer that, we need to first specify the knowledge and observation models as well as the cost (reward) functions for individual agents. Then we will derive the optimal choices of agents in a distributed manner. The resulting system dynamics correspond to a Nash Equilibrium of the system.

\subsection{Knowledge and Observation Model}
A knowledge and observation model for agent $i$ includes two aspects: what does agent $i$ know about itself, and what does agent $i$ know about others? The knowledge about any agent $j$ includes the dynamic function of agent $j$ and the cost function of agent $j$. The observation corresponds to run-time measurements, i.e., the observation of any agent $j$ includes the run-time state $x_{j,k}$ and the run-time control $u_{j,k}$. In the following discussion, regarding the knowledge and observation model, we make the following assumptions:
\begin{itemize}
\item An agent knows its own dynamics and cost function;
\item All agents are homogeneous in the sense that they share the same dynamics and cost functions. And agents know that all agents are homogeneous, hence they know others' dynamics and cost functions;\sidenote{Not knowing other agents' dynamics or cost functions will result in information asymmetry, which creates difficulty in the analysis. Nonetheless, the assumption can be relaxed in the future.}
\item At time $k$, agents can measure $x_{j,k}$ for all $j$. But they cannot measure $u_{j,k}$ until time $k+1$. Hence, the agents are playing a simultaneous game. They need to infer others' decisions when making their own decisions at any time $k$.
\end{itemize}

\subsection{Cost Function}

We consider two conflicting interests for every agent:\sidenote{The identification of these two conflicting interests is purely empirical. To build realistic cost functions, we need to either study the real world data or conduct human subject experiments.}
\begin{itemize}
\item Limit the activity level to minimize the chance to get infected;
\item Maintain a certain activity level for living.
\end{itemize}
We define the run-time cost for agent $i$ at time $k$ as
\begin{equation}
l_{i,k} = x_{i,k+1} + \alpha_i p(u_{i,k}),
\end{equation}
where $x_{i,k+1}$ corresponds to the first interest, $p(u_{i,k})$ corresponds to the second interest,  and $\alpha_i > 0$ adjusts the preference between the two interests. The function $p(u)$ is assumed to be smooth.\sidenote{The function $p(u)$ can be a decreasing function on $[0,1]$, meaning that the higher the activity level, the better. The function $p(u)$ can also be a convex parabolic function on $[0,1]$ with the minimum attained at some $u^*$, meaning that the activity level should be maintained around $u^*$.} Due to our homogeneity assumption on agents, they should have identical preferences, i.e., $\alpha_i = \alpha$ for all $i$. 

Agent $i$ chooses its action at time $k$ by minimizing the expected cumulative cost in the future:
\begin{equation}\label{eq: repeated game}
u_{i,k} = \arg\min \mathbb{E}[\sum_{t=k}^{\infty} \gamma^{t-k} l_{i,k} ],
\end{equation}
where $\gamma\in[0,1]$ is a discount factor. The objective function depends on all agents' current and future actions. It is difficult to directly obtain an analytical solution of \eqref{eq: repeated game}. Later we will use multi-agent reinforcement learning to obtain a numerical solution.

In this section, to simplify the problem, we consider a single stage game\sidenote{The formulation \eqref{eq: repeated game} corresponds to a repeated game as opposed to the single stage game. Repeated games capture the idea that an agent will have to take into account the impact of its current action on the future actions of others. This impact is called the agent's reputation. The interaction is more complex in a repeated game than that in a single stage game.} where the agents have zero discount of the future, i.e., $\gamma = 0$. Hence the objective function is reduced to
\begin{equation}
u_{i,k} = \arg\min \mathbb{E}[l_{i,k} ],
\end{equation}
which only depends on the current actions of agents. According to the state transition probability in \cref{table:dynamics}, the expected cost is
\begin{equation}\label{eq: expected cost}
\mathbb{E}[l_{i,k} ]  = \left\{\begin{array}{ll}
1-\Pi_{j\in\mathcal{I}_k} (1-\min\{u_i,u_j\}) + \alpha_i p(u_{i,k}) & \text{if }x_{i,k} = 0\\
1 + \alpha_i p(u_{i,k}) & \text{if }x_{i,k} = 1
\end{array}\right..
\end{equation}

\subsection{Nash Equilibrium}
\begin{marginfigure}
\begin{center}
\includegraphics[width=\linewidth]{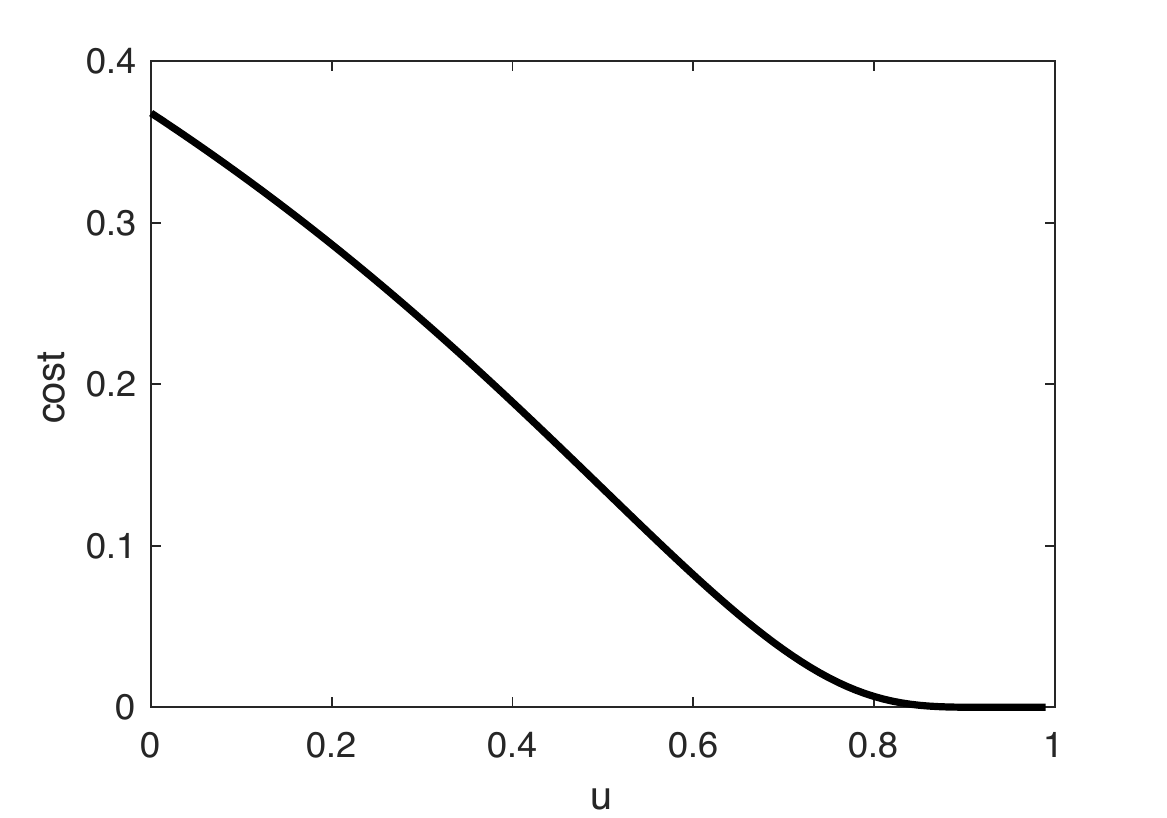}
\caption{Illustration of the curve $\exp(\frac{1}{u-1})$.}
\label{fig: p}
\end{center}
\end{marginfigure}
According to \eqref{eq: expected cost}, the expect cost for an infected agent only depends on its own action. Hence the optimal choice for an infected agent is $u_{i,k} = \bar u := \arg\min_u p(u)$. Then the optimal choice for a healthy agent satisfies:
\begin{eqnarray}
u_{i,k} &=& \arg\min_u [1-\Pi_{j\in\mathcal{I}_k} (1-\min\{u,\bar u\}) + \alpha_i p(u)],\\
&=&\arg\min_u [1-(1-\min\{u,\bar u\})^{m_k} + \alpha_i p(u)].\label{eq: healthy agent cost}
\end{eqnarray}
Note that the term $1-(1-\min\{u,\bar u\})^{m_k}$ is positive and is increasing for $u\in[0,\bar u]$ and then constant for $u\in[\bar u, 1]$. Hence, the optimal solution for \eqref{eq: healthy agent cost} should be smaller than $\bar u = \arg\min_u p(u)$.\sidenote{If $u\geq\bar u$, then \eqref{eq: healthy agent cost} becomes $\arg\min_u [1-(1- \bar u)^{m_k} + \alpha_i p(u)]$, whose optimal solution is $u = \bar u$ with cost $[1-(1- \bar u)^{m_k} + \alpha_i p(\bar u)]$. If $u\leq\bar u$, then \eqref{eq: healthy agent cost} becomes $\arg\min_u J(u)$ where $J(u) = [1-(1- u)^{m_k} + \alpha_i p(u)]$. Since $\frac{\partial}{\partial u}\mid_{\bar u} J(u) > 0$, the optimal solution satisfies that $u< \bar u$ with cost $J(u) < J(\bar u)$. Note that $J(\bar u)$ equals to the smallest cost for the case $u\leq\bar u$. Hence the optimal solution for \eqref{eq: healthy agent cost} satisfies that $u< \bar u$.} 
Then the objective in \eqref{eq: healthy agent cost} can be simplified as $1-(1-u)^{m_k} + \alpha_i p(u)$. 
In summary, the optimal actions for both the infected and the healthy agents in the Nash Equilibrium can be compactly written as
\begin{equation}\label{eq: Nash}
u_{i,k} = \arg\min_u \{1-(1-u)^{m_k}(1-x_{i,k}) + \alpha_i p(u)\}, \forall i.
\end{equation}

\paragraph{Example} 
Consider the previous example with four agents shown in \cref{fig:example}. Define 
\begin{equation}\label{eq: cost on activity}
p(u) = \exp(\frac{1}{u-1}),
\end{equation}
which is a monotonically decreasing function as illustrated in \cref{fig: p}. Then the optimal actions in the Nash Equilibrium for this specific problem satisfy:
\begin{equation}\label{eq: example Nash}
u_{i,k} = \arg\min_u \{u+x_{i,k}-ux_{i,k} + \alpha_i \exp(\frac{1}{u-1})\}, \forall i.
\end{equation} 
\begin{marginfigure}
\begin{center}
\includegraphics[width=\linewidth]{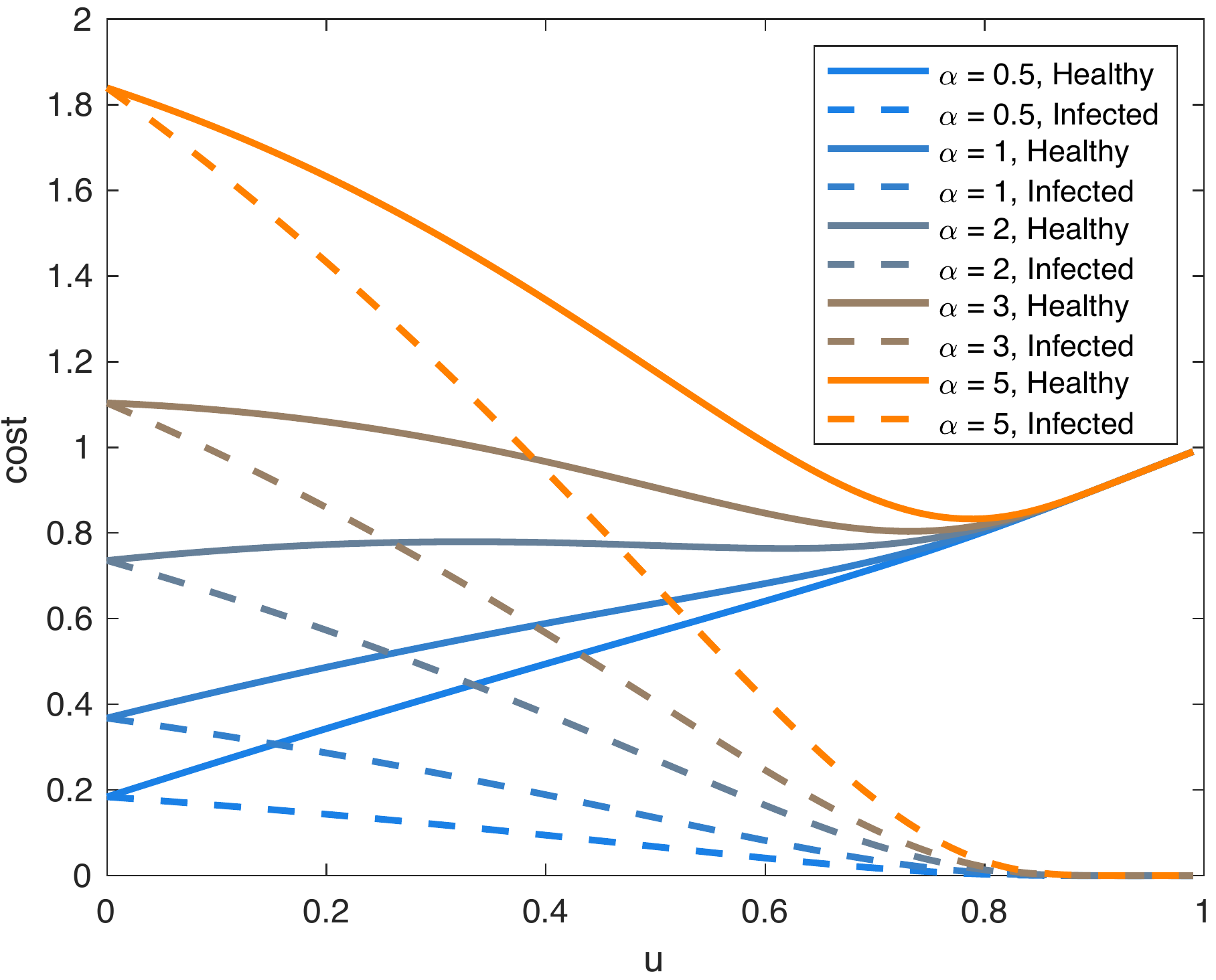}
\caption{Illustration of the objective function in \eqref{eq: example Nash} under different conditions.}
\label{fig: alpha}
\end{center}
\end{marginfigure}
Solving for \eqref{eq: example Nash}, for infected agents, $u_{i,k} = 1$. For healthy agents, the choice also depends on $\alpha_i$ as illustrated in \cref{fig: alpha}. We have assumed that $\alpha_i = \alpha$ which is identical for all agents. We further assume that $\alpha < 2$ such that the optimal solution for healthy agents should be $u_{i,k} = 0$. The optimal actions and the corresponding costs for all agents are listed in \cref{fig: example cost}. In the Nash Equilibrium, no agent will meet each other, since all agents except agent $1$ reduce their activity levels to zero. The actual cost (received at the next time step) equals to the expected cost (computed at the current time step).
\begin{table}[h]
\caption{List of the agent decisions and associated costs in the Nash Equilibrium in the four-agent example.}
\begin{center}\small
\begin{tabular}{c|cccc}
Agent ID & State $x_{i,k}$ & Optimal $u_{i,k}$ & Optimal $\mathbb{E}[l_{i,k}]$ & Actual $l_{i,k}$\\
\hline
1 & 1 & 1 & 1 & 1\\
2,3,4 & 0 & 0 & $\alpha \exp(-1)$ & $\alpha \exp(-1)$\\
\hline
Total & & & & $1+ 3\alpha \exp(-1)$
\end{tabular}
\end{center}
\label{fig: example cost}
\end{table}%

However, let us consider another situation where the infected agent chooses $0$ activity level and all other healthy agents choose $1$ activity level. The resulting costs are summarized in \cref{fig: example cost global}.  Obviously, the overall cost is reduced in the new situation. However, this better situation cannot be attained spontaneously by the agents, due to externality of the system which will be explained below.
\begin{table}[h]
\caption{List of the agent decisions and associated costs in a situation better than the Nash Equilibrium in the four-agent example.}
\begin{center}\small
\begin{tabular}{c|cccc}
Agent ID & State $x_{i,k}$ & Optimal $u_{i,k}$ & Optimal $\mathbb{E}[l_{i,k}]$ & Actual $l_{i,k}$\\
\hline
1 & 1 & 0 & 1+$\alpha \exp(-1)$ & 1+$\alpha \exp(-1)$\\
2,3,4 & 0 & 1 & 0 & 0\\
\hline
Total & & & & $1+ \alpha \exp(-1)$
\end{tabular}
\end{center}
\label{fig: example cost global}
\end{table}%

\subsection{Dealing with Externality}
For a multi-agent system, define the system cost as a summation of the individual costs:
\begin{equation}
L_k := \sum_i l_{i,k}.
\end{equation}
The system cost in the Nash Equilibrium is denoted $L^*_k$, which corresponds to the evaluation of $L_k$ under agent actions specified in \eqref{eq: Nash}. On the other hand, the optimal system cost is defined as
\begin{equation}\label{eq: system optimum}
L^o_k := \min_{u_{i,k},\forall i} L_k.
\end{equation}
The optimization problem \eqref{eq: system optimum} is solved in a centralized manner, which is different from how the Nash Equilibrium is obtained. To obtain the Nash Equilibrium, all agents are solving their own optimization problems independently. Although their objective functions depend on other agents' actions, they are not jointly make the decisions, but only ``infer" what others will do. 
By definition, $L^o_k \leq L^*_k$. In the example above, $L^*_k = 1+3\alpha \exp(-1)$ and $L^o_k = 1+\alpha \exp(-1)$.
The difference $L^*_k - L^o_k$ is called the \textbf{loss of social welfare}.
In the epidemic model, the loss of social welfare is due to the fact that bad consequences (i.e., infecting others) are not penalized in the cost functions of the infected agents. Those unpenalized consequences are called \textit{externality}. There can be both positive externality and negative externality. Under positive externality, agents  are lacking motivations to do things that are good for the society. Under negative externality, agents are lacking motivations to prevent things that are bad for the society. In the epidemic model, there are negative externality with infected agents.

To improve social welfare, we need to ``internalize" externality, i.e., add penalty for ``spreading" the disease. Now let us redefine agent $i$'s run-time cost as
\begin{equation}\label{eq: shaped cost}
\tilde l_{i,k} = x_{i,k+1} + \alpha_i p(u_{i,k}) + x_{i,k} q(u_{i,k}),
\end{equation}
where $q(\cdot)$ is a monotonically increasing function.
The last term $x_{i,k}q(u_{i,k})$ does not affect healthy agents since $x_{i,k} = 0$, but adds a penalty for infected agents if they choose large activity level. One candidate function for $q(u)$ is $1-(1-u)^{m_k}$. In the real world, such ``cost shaping" using $q$ can be achieved through social norms or government regulation. The expected cost becomes
\begin{equation}\label{eq: expected cost tilde}
\mathbb{E}[\tilde l_{i,k} ]  = \left\{\begin{array}{ll}
1-\Pi_{j\in\mathcal{I}_k} (1-\min\{u_i,u_j\}) + \alpha_i p(u_{i,k}) & \text{if }x_{i,k} = 0\\
1 + \alpha_i p(u_{i,k}) + q(u_{i,k}) & \text{if }x_{i,k} = 1
\end{array}\right.
\end{equation}
Suppose the function $q$ is well tuned such that the $\arg\min_u [ \alpha_i p(u) + q(u)] = 0$. Then although the expected costs for infected agents are still independent from others, their decision is considerate to healthy agents. When the infected agents choose $u =0$, then for healthy agents, the expected cost becomes $\alpha_i p(u_{i,k})$, meaning that they do not need to worry about getting infected.
Let us now compute the resulting Nash Equilibrium under the shaped costs using the previous example.

\paragraph{Example}
In the four-agent example, set $q(u) = u$. Then $\arg\min_u [ \alpha p(u) + u] = 0$. Hence agent 1 will choose $u_{1,k} = 0$. For agents $i= 2,3,4$, they will choose $u_{i,k}=1$ since they are only minimizing $p(u)$. The resulting costs are summarized in \cref{fig: example cost improved}. With the shaped costs, the system enters into a better Nash Equilibrium which indeed aligns with the system optimum in \eqref{eq: system optimum}. A few remarks:
\begin{itemize}
\item Cost shaping did not increase the overall cost for the multi-agent system.
\item The system optimum remains the same before and after cost shaping.
\item Cost shaping helped agents to arrive at the system optimum without centralized optimization.
\end{itemize}

\begin{table}[h]
\caption{List of the agent decisions and associated costs in the Nash Equilibrium with shaped cost functions in the four-agent example.}
\begin{center}\small
\begin{tabular}{c|cccc}
Agent ID & State $x_{i,k}$ & Optimal $u_{i,k}$ & Optimal $\mathbb{E}[\tilde l_{i,k}]$ & Actual $\tilde l_{i,k}$\\
\hline
1 & 1 & 0 & 1+$\alpha \exp(-1)$ & 1+$\alpha \exp(-1)$\\
2,3,4 & 0 & 1 & 0 & 0\\
\hline
Total & & & & $1+ \alpha \exp(-1)$
\end{tabular}
\end{center}
\label{fig: example cost improved}
\end{table}%

\section{Multi-Agent Reinforcement Learning}
We have shown how to compute the Nash Equilibrium of the multi-agent epidemic model in a single stage. However, it is analytically intractable to compute the Nash Equilibrium when we consider repeated games \eqref{eq: repeated game}. The complexity will further grow when the number of agents increases and when there are information asymmetry. Nonetheless, we can apply multi-agent reinforcement learning\cite{bucsoniu2010multi} to numerically compute the Nash Equilibrium. Then the evolution of the pandemic can be predicted by simulating the system under the Nash Equilibrium. 

\subsection{Q Learning}
As evident from \eqref{eq: Nash}, the optimal action for agent $i$ at time $k$ is a function of $x_{i,k}$ and $m_k$. Hence we can define a Q function (action value function) for agent $i$ as
\begin{equation}
Q_i: x_{i,k}\times m_k\times u_{i,k}\mapsto \mathbb{R}.
\end{equation}
According to the assumptions made in the observation model, all agents can observe $m_k$ at time $k$. 
For a single stage game, we have derived in \eqref{eq: Nash} that $Q_i (x,m,u)= 1-(1-u)^{m}(1-x) + \alpha_i p(u)$. For repeated games \eqref{eq: repeated game}, we can learn the Q function using temporal different learning. At every time $k$, agent $i$ chooses its action as
\begin{equation}\label{eq: optimal action}
u_{i,k} = \arg\min_u Q_i(x_{i,k}, m_k, u).
\end{equation}
After taking the action $u_{i,k}$, agent $i$ observes $x_{i,k+1}$ and $m_{k+1}$ and receives the cost $l_{i,k}$ at time $k+1$. Then agent $i$ updates its Q function:
\begin{eqnarray}
&Q_i (x_{i,k},m_k,u_{i,k}) \leftarrow Q_i (x_{i,k},m_k,u_{i,k}) + \eta \delta_{i,k},\\
&\delta_{i,k} = l_{i,k} + \gamma \min_u Q_i(x_{i,k+1},m_{k+1},u) - Q_i(x_{i,k},m_k,u_{i,k}),
\end{eqnarray}
where $\eta$ is the learning gain and $\delta_{i,k}$ is the temporal difference error.

All agents can run the above algorithm to learn their $Q$ functions during the interaction with others. However, the algorithm introduced above has several problems:
\begin{itemize}
\item Exploration and limited rationality.

There is no exploration in \eqref{eq: optimal action}. Indeed, Q-learning is usually applied together with $\epsilon$-greedy where with probability $1-\epsilon$, the action $u_{i,k}$ is chosen to be the optimal action in \eqref{eq: optimal action}, and with probability $\epsilon$, the action is randomly chosen with a uniform distribution over the action space. The $\epsilon$-greedy approach is introduced mainly from an algorithmic perspective to improve convergence of the learning process. When applied to the epidemic model, it has a unique societal implication. When agents are randomly choosing their behaviors, it represents the fact that agents have only limited rationality. Hence in the learning process, we apply $\epsilon$-greedy as a way to incorporate exploration for faster convergence as well as to take into account  limited rationality of agents.

\item Data efficiency and parameter sharing.

Keeping separated Q functions for individual agents is not data efficient. An agent may not be able to collect enough samples to properly learn the desired Q function. Due to the homogeneity assumptions we made earlier about agents' cost functions, it is more data efficient to share the Q function for all agents. Its societal implication is that agents are sharing information and knowledge with each other. Hence, we apply parameter sharing\cite{gupta2017cooperative} as a way to improve data efficiency as well as to consider information sharing among agents during the learning process.\sidenote{In a more complex situation where agents are not homogeneous, it is desired to have parameter sharing with a smaller group of agents, instead of parameter sharing will all agents.}
\end{itemize}

With the above modifications, the multi-agent Q learning algorithm\cite{hu2003nash} is summarized below.
\begin{itemize}
\item For every time step $k$, agents choose their actions as:
\begin{equation}
u_{i,k} = \left\{\begin{array}{ll}
\arg\min_u Q(x_{i,k}, m_k, u) & \text{probability }1-\epsilon\\
\text{random} &\text{probability }\epsilon \end{array}\right. \forall i.
\end{equation}

\item At the next time step $k+1$, agents observe the new states $x_{i,k+1}$ and receive rewards $l_{i,k}$ for all $i$. Then the Q function is updated:
\begin{eqnarray}
&Q (x_{i,k},m_k,u_{i,k}) \leftarrow Q (x_{i,k},m_k,u_{i,k}) + \eta \delta_{i,k},\forall i,\\
&\delta_{i,k} = l_{i,k} + \gamma \min_u Q(x_{i,k+1},m_{k+1},u) - Q(x_{i,k},m_k,u_{i,k}).
\end{eqnarray}

\end{itemize}

\begin{figure}[htbp]
\begin{center}
\subfloat[Case 1 with discount $\gamma = 0$ and the original runtime cost.\label{fig: marl-1}]{
\includegraphics[width=0.5\linewidth]{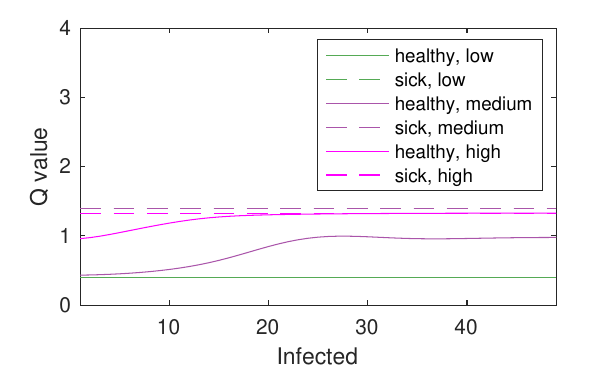}
\includegraphics[width=0.48\linewidth]{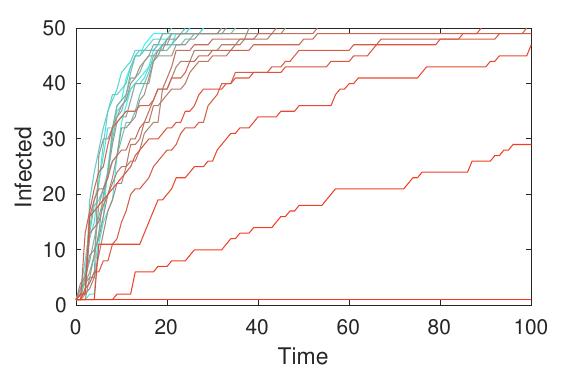}}

\subfloat[Case 2 with discount $\gamma = 0.5$ and the original runtime cost.\label{fig: marl-2}]{
\includegraphics[width=0.5\linewidth]{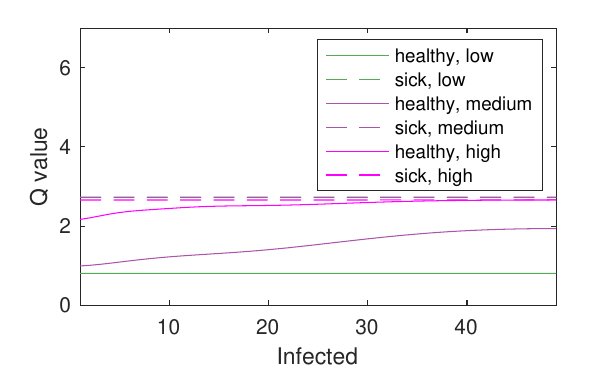}
\includegraphics[width=0.48\linewidth]{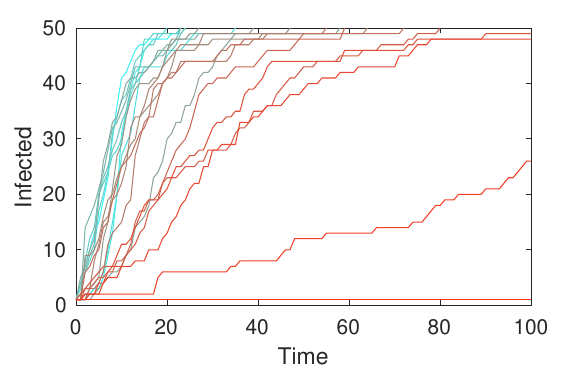}}

\subfloat[Case 3 with discount $\gamma = 0.5$ and the shaped runtime cost.\label{fig: marl-3}]{
\includegraphics[width=0.5\linewidth]{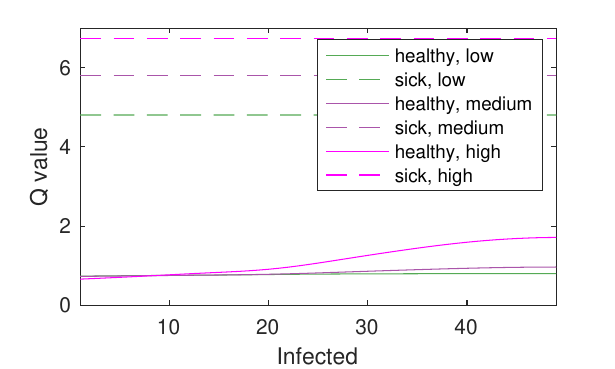}
\includegraphics[width=0.48\linewidth]{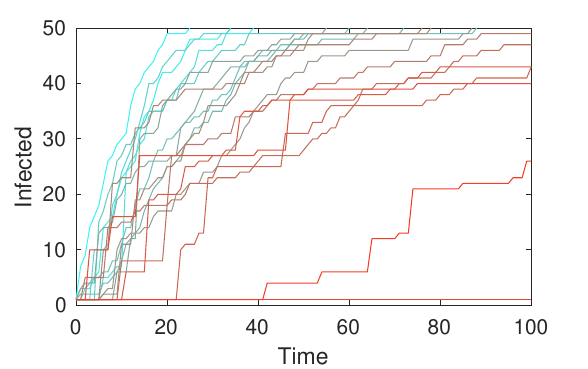}}

\caption{Results in the multi-agent Q learning under the microscopic epidemic model. The left plots show the learned Q values after 200 episodes. The horizontal axis corresponds to $m_k$. The vertical axis corresponds to the Q values. Solid curves are for healthy agents $x_{i,k}=0$. Dashed curves are for infected agents $x_{i,k} = 1$. Green curves are for low activity levels $u_{i,k} = 0$. Purple curves are for medium activity levels $u_{i,k} = 1/M$. Magenta curves are for high activity levels $u_{i,k} = 10/M$. The right plots illustrate the system trajectories for episodes $10, 20,\ldots, 200$, blue for earlier episodes and red for later episodes. In the last episode in all cases where there is no exploration $\epsilon=0$, the system trajectories are horizontal with $m_k \equiv 1$. }
\label{fig: marl}
\end{center}
\end{figure}

\paragraph{Example}
In this example, we consider $M=50$ agents in the system. Only one agent is infected in the beginning. The run-time cost is the same as in the example in the \textit{distributed optimal control} section, i.e., $l_{i,k} = x_{i,k+1} + \alpha \exp (\frac{1}{u_{i,k}-1})$ where $\alpha$ is chosen to be $1$. For simplicity, the action space is discretized to be $\{0, 1/M, 10/M\}$, called as low, medium, and high. Hence the Q function can be stored as a $2\times M\times 3$ matrix. In the learning algorithm, the learning rate is set to $\eta = 1$. The exploration rate is set to decay in different episodes, i.e., $\epsilon = 0.5(1-E/\max E)$ where $E$ denotes the current episode and the maximum episode is $\max E = 200$. The Q function is initialized to be $10$ for all entries. Three different cases are considered. For each case, we illustrate the Q function learned after 200 episodes as well as the system trajectories for episodes $10, 20,\ldots, 200$, blue for earlier episodes and red for later episodes. The results are shown in \cref{fig: marl}.
\begin{itemize}
\item Case 1: discount $\gamma = 0$ with runtime cost $l_{i,k}$. 

With $\gamma = 0$, this case reduces to a single stage game as discussed in the \textit{distributed optimal control} section. The result should align with the analytical Nash Equilibrium in \eqref{eq: Nash}. As shown in the left plot in \cref{fig: marl}(a), the optimal action for a healthy agent is always \textit{low} (solid green), while the optimal action for an infected agent is always \textit{high} (dashed magenta). The Q values for infected agents do not depend on $m_k$. The Q values for healthy agents increase when $m_k$ increases if the activity level is not zero,  due to the fact that: for a fixed activity level, the chance to get infected is higher when there are more infected agents in the system. All these results align with our previous theoretical analysis. Moreover, as shown in the right plot in \cref{fig: marl}(a), the agents are learning to flatten the curve across different episodes. 
\item Case 2: discount $\gamma = 0.5$ with runtime cost $l_{i,k}$. 

Since the agents are now computing cumulative costs as in \eqref{eq: repeated game}, the corresponding Q values are higher than those in case 1. However, the optimal actions remain the same, \textit{low} (solid green) for healthy agents, \textit{high} (dashed magenta) for infected agents, as shown in the left plot in \cref{fig: marl}(b). The trends of the Q curves also remain the same: the Q values do not depend on $m_k$ for infected agents and for healthy agents whose activity levels are zero. However, as shown in the right plot in \cref{fig: marl}(b), the agents learned to flatten the curve faster than in case 1, mainly because healthy agents are more cautious (converge faster to low activity levels) when they start to consider cumulative costs.

\item Case 3: discount $\gamma = 0.5$ with shaped runtime cost $\tilde l_{i,k}$ in \eqref{eq: shaped cost}.

The shaped cost changes the optimal actions for all agents as well as the resulting Q values. 
As shown in the left plot in \cref{fig: marl}(c), the optimal action for an infected agent is \textit{low} (dashed green), while that for a healthy agent is \textit{high} (solid magenta) when $m_k$ is small and \textit{low} (solid green) when $m_k$ is big. Note that when $m_k$ is high, the healthy agents still prefer low activity level, though the optimal actions for infected agents are low. That is because: due to the randomization introduced in $\epsilon$-greedy, there is still chance for infected agents to have medium or high activity levels. When $m_k$ is high, the healthy agents would rather limit their own activity levels to avoid the risk to meet with infected agents that are taking random actions. This result captures the fact that agents understand others may have limited rationality and prefer more conservative behaviors.
We observe the same trends for the Q curves as the previous two cases: the Q values do not depend on $m_k$ for infected agents and for healthy agents whose activity levels are not zero. In terms of absolute values, the Q values for infected agents are higher than those in case 2 due to the additional cost $q(u)$ in $\tilde l_{i,k}$. The Q values for healthy agents are smaller than those in case 2 for medium and high activity levels, since the chance to get infected is smaller as infected agents now prefer low activity levels. The Q values remain the same for healthy agents with zero activity levels. 
With shaped costs, the agents learned to flatten the curve even faster than in case 2, as shown in the right plot in \cref{fig: marl}(c), since the shaped cost encourages infected agents to lower their activity levels. 
\end{itemize}

\section{Discussion and Future Work}
\paragraph{Agents vs humans}
The epidemic model can be used to analyze real-world societal problems.
Nonetheless, it is important to understand the differences between agents and humans.
We can directly design and shape the cost function for agents, but not for humans. 
For agents, their behavior is predictable once we fully specify the problem (i.e., cost, dynamics, measurement, etc). Hence we can optimize the design (i.e., the cost function) to get desired system trajectory.
For humans, their behavior is not fully predictable due to limited rationality. We need to constantly modify the knowledge and observation model as well as the cost function to match the true human behavior.

\paragraph{Future work}
The proposed model is in its preliminary form. Many future directions can be pursued.
\begin{itemize}
\item Relaxation of assumptions.

We may add more agent states to consider recovery, incubation period, and death.  We may consider the fact that the interaction patterns among agents are not uniform. We may consider a wide variety of agents who are not homogeneous. For example, health providers and equipment suppliers are key parts in fighting the disease. They should receive lower cost (higher reward) for maintaining or even expanding their activity levels than ordinary people. Their services can then lead to higher recovery rate. In addition, we may relax the assumptions on agents' knowledge and observation models, to consider information asymmetry as well as partial observation. For example, agents cannot get immediate measurement whether they are infected or not, or how many agents are infected in the system.

\item Realistic cost functions for agents.

The cost functions for agents are currently hand-tuned. We may learn those cost functions from data through inverse reinforcement learning. Those cost functions can vary for agents from different countries, different age groups, and different occupations. Moreover, the cost functions carry important cultural, demographical, economical, and political information. A realistic cost function can help us understand why we observe significantly different outcomes of the pandemic around the world, as well as enable more realistic predictions into the future by fully considering those cultural, demographical, economical, and political factors.

\item Incorporation of public policies.

For now, the only external intervention we introduced is cost shaping. We may consider a wider range of public policies that can change the closed-loop system dynamics. For example, shut-down of transportation, isolation of infected agents, contact tracing, antibody testing, etc.

\item Transient vs steady state system behaviors.

We have focused on the steady state system behaviors in the Nash Equilibrium. However, as agents live in a highly dynamic world, it is not guaranteed that a Nash Equilibrium can always be attained. While agents are learning to deal with unforeseen situations, there are many interesting transient dynamics, some of which is captured in \cref{fig: marl}, i.e., agents may learn to flatten the curve at different rates.  
Methods to understand and predict transient dynamics may be developed in the future.

\item Validation against real world historical data.

To use the proposed model for prediction in the real world, we need to validate its fidelity again the historical data. The validation can be performed on the $m_k$ trajectories, i.e., for the same initial condition, the predicted $m_k$ trajectories should align with the ground truth $m_k$ trajectories. 
\end{itemize}

\section{Conclusion}
This paper introduced a microscopic multi-agent epidemic model, which explicitly considered the consequences of individual's decisions on the spread of the disease. In the model, every agent can choose its activity level to minimize its cost function consisting of two conflicting components: staying healthy by limiting activities and maintaining high activity levels for living. We solved for the optimal decisions for individual agents in the framework of game theory and multi-agent reinforcement learning. Given the optimal decisions of all agents, we can make predictions about the spread of the disease. The system had negative externality in the sense that  infected agents did not have enough incentives to protect others, which then required external interventions such as cost shaping. 
We identified future directions were pointed out to make the model more realistic.

\bibliography{main}
\bibliographystyle{plainnat}

\end{document}